\documentstyle[aps,twocolumn,epsf]{revtex}

\begin{document} 
\draft
\twocolumn[\hsize\textwidth\columnwidth\hsize\csname
@twocolumnfalse\endcsname
\tightenlines
\title{DNA-condensation, redissolution and mesocrystals induced by tetravalent counterions} 
\author{E.Allahyarov $^{1,2,3}$,\,\, H.L\"owen $^{1}$,\,\, G.Gompper $^{2}$}
\address{{1} Institut\, f\"{u}r\, Theoretische \,Physik
  II,\,Heinrich-Heine-Universit\"{a}t\,
 D\"{u}sseldorf,\,\mbox{D-40225}\,D\"{u}sseldorf, \,Germany}
\address{{2} Institut\, f\"ur\, Festk\"orperforschung, Forschungszentrum 
J\"ulich, \,\mbox{D-52425} \, J\"ulich, Germany }
\address{{3} Institute for High Temperatures, Russian Academy of
Sciences, Izhorskaya street 13/19, \mbox{127412} Moscow, Russia}
\maketitle
\begin{abstract}
 The distance-resolved effective interaction potential 
 between two parallel DNA molecules is calculated by computer 
simulations with explicit tetravalent 
counterions and monovalent salt. Adding counterions first yields
an attractive minimum in the potential at short distances which then disappears
in favor of a shallower  minimum at larger separations.
The resulting phase diagram includes a  DNA-condensation and  
redissolution transition and a stable mesocrystal with an intermediate lattice
constant for high counterion concentration.
\end{abstract}
]
\renewcommand{\thepage}{\hskip 8.9cm \arabic{page} \hfill Typeset
using REV\TeX }
\narrowtext
Multivalent polyamines such as trivalent spermidine (Spd) and
tetravalent spermine (Spe) are abundant in living cells and play a key role in
maintaining cellular DNA in a compact state \cite{bloomfield1997,cohen1998book,saminathan}. 
They modulate ion channel activities of cells and are essential for
normal cell growth. 
 Polyamines also facilitate
the packaging of DNA in certain viruses. The implications can effectively be applied
in gene delivery and in the field of genetic therapy.  Under physiologic ionic and pH
conditions, the polyamines are positively charged and hence DNA
is their prime target of interaction.
 \hskip -1cm
\vspace{-0.3cm}
\begin{figure}
   \epsfxsize=8cm 
   \epsfysize=8cm 
~\hfill\epsfbox{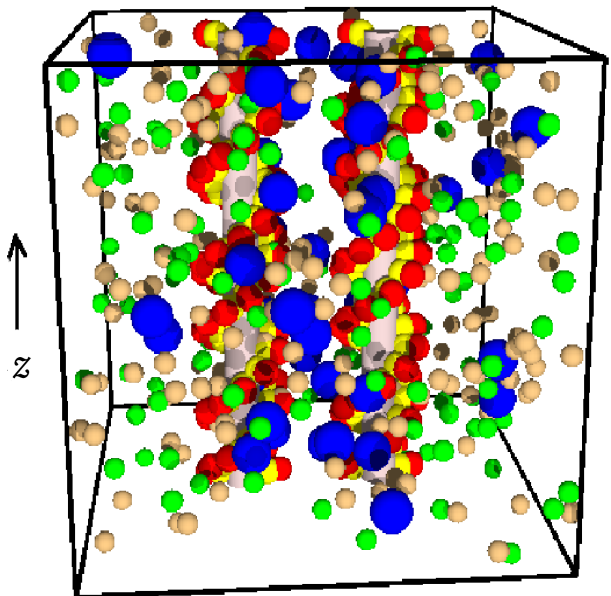}\hfill~  
 \caption{(Color online) Typical snapshot in the simulation cell.
The DNA molecules are shown as two parallel in $\it z$ direction rods over-wrapped by two
strings of light grey (neutral sphere in MAM (see text), colored yellow in online
figure) and grey (phosphate sphere in MAM,  red in online
figure) spheres. The tetravalent Spe ions are
   shown as big black (blue in online figure) spheres. Light grey
   (yellow in online figure) spheres represent
   coions, and dark grey (green in online figure) spheres are monovalent counterions.}
 \label{figure_2}
\end{figure}
 In the last decade different experiments have shown a condensation
 and a subsequent redissolution 
of DNA for increasing polyamine concentration $C$
 \cite{saminathan,pelta1996,longtriplex1999saminathan,solis2001reentrance,solis2002,vries2001recondensation,murayama2003reentrance}. The condensation and redissolution
occur at concentrations $C_c$ and $C_d$ respectively. Between these two thresholds,
for $C_c < C < C_d$, there is a coexistence of a liquid-like dense DNA phase
and a very dilute DNA solution. Several theoretical explanations have been presented for
the condensation based on counterion-induced attractions between the
DNA molecules
 \cite{bloomfield1997,saminathan,raspaud1999,nguen_reentrant,bloomfield1996cosb,burak2003,arscott1990dnacollapse,Solis2000}. 
However, the origin of the experimentally found redissolution transition 
\cite{raspaud1999,trubetskoy2003reentrance} is 
not understood on a molecular level.
Proposed mechanisms range from an increased DNA-hydrophilicity induced by polyamine bindings
\cite{longtriplex1999saminathan} to Bjerrum pair condensation 
\cite{solis2001reentrance,solis2002,tanaka2001chargeinversion}
formed by multivalent counter- and monovalent coions and DNA-overcharging \cite{nguen_reentrant}.

In this letter we investigate the condensation
and redissolution of DNA on a molecular level by using
primitive-model computer simulations with explicit tetravalent counterions
and monovalent salt ions.
We trace back the condensation and redissolution to the  attractions
in the distance-dependent effective potential $U(R)$ between two parallel DNA molecules
with $R$ denoting the radial distance between their two centers.
In fact, the depth and position of the attractive minimum play a crucial
role in whether there is liquid-gas-like phase separation. For
monovalent microions, the interaction is repulsive \cite{ourfirstDNApaper}.
For increasing tetravalent counterion concentration $C$,
an attractive minimum in $U(R)$ at small separations $R\approx 28 \AA$
shows up which  then disappears
in favor of a shallower  minimum at larger separations $R\approx 39 \AA$.
Using two-dimensional liquid-state theory for the fluid and lattice sums for
the solid phases, we calculate the phase diagram for columnar DNA assemblies.
It  includes the  DNA-condensation and 
redissolution transition  and the associated threshold concentrations $C_c$ and $C_d$
are in  agreement with the experimental data. For high
concentrations $C \gtrsim 160$mM, we 
 predict a stable hexagonal mesocrystal with an intermediate lattice
constant which can coexist either with a dense hexagonal crystal or a dilute solution.

In our computer simulations, we consider B-DNA molecules 
which form  a  double helix with a
pitch length  of $P$=34\AA \,
 and $N_P$=20 phosphate charges per pitch
 using the realistic groove-geometry and charge pattern of the 
 Montoro-Abascal model (MAM)
 \cite{montoro1998,oursecondDNApaper}.
A single DNA molecule or a pair of parallel DNA molecules, which are
oriented in the {\it z} direction, are placed 
on the {\it xy} diagonal of a cubic simulation
box of length $L=102\AA$ which is three times the pitch length. 
The box also contains $N_Q$ tetravalent ions, $N_- = N_{s} +  4N_Q$
monovalent coions and $ N_+ = N_{s} + N_p$ monovalent counterions \cite{footnote}. Here 
$N_s$ is the number of added salt ion pairs, and 
$N_p$ is fixed by the
DNA charge due to the constraint
of global charge neutrality; $N_p=3N_P=60$ for a single
DNA and $N_p=6N_P=120$ for two DNA molecules in the box. 
All ions are modelled as charged hard spheres with
$d_Q =8\AA$ denoting the diameter of the tetravalent counterions.
All diameters of the  monovalent microions are assumed to be equal
and we choose them to be $d_c$=4\AA.
Hereafter we shall call the tetravalent counterions spermine (Spe) since
experimental data support the idea that it is the charge of
a counterion, rather than its structural specificities, which is important in
DNA condensation and recondensation processes.

Periodic boundary conditions in all three
directions are applied.  The whole system is held at room
temperature $T=298K$ and the water is modelled as a continuous
dielectric medium with $\epsilon=80$. 
The interaction potentials between
the different particle species 
 are a combination of hard core and Coulomb potentials.
We have performed extensive grand canonical molecular dynamics (GCMD)
 simulations, as described  in
 Ref.~\cite{ourthirdDNApaper}, for a
 range of different Spe and salt concentrations.
A typical configurational snapshot in the 
simulation box is shown in Figure~\ref{figure_2}.

\begin{figure}
   \epsfxsize=8cm
   \epsfysize=8cm 
~\hfill\epsfbox{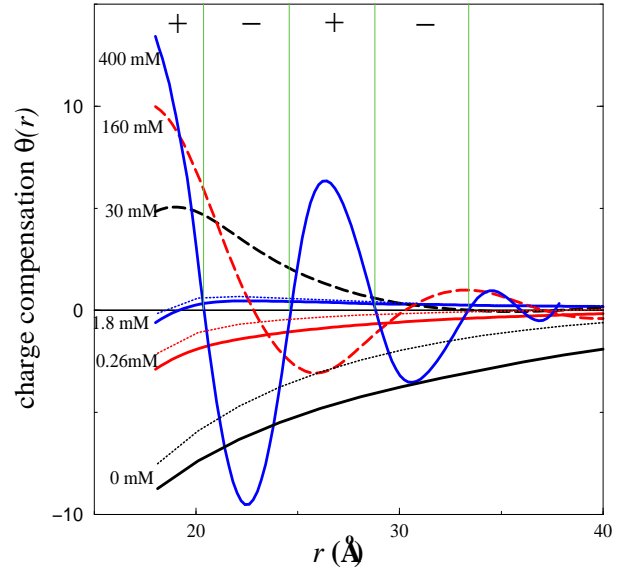}\hfill~ 
  \caption{(Color online) The DNA overcharging parameter $\theta$ versus
   the distance $r$ from the DNA axis for different Spe concentrations. 
 Thick lines- $c_s$ = 25mM, 
  thin lines- $c_s$ =100mM. 
 The Spe concentrations are shown next to
  corresponding curves. Vertical lines, serving as guide for eye,
   separate the charge layers of
   different sign for $C$=400mM.
} \label{figure_3}
\end{figure}

First we consider a {\it single} DNA molecule in the simulation box. It is
known that, in the presence of multivalent ions, the ionic cloud may not only
compensate the polyion charge, but even exceed it, resulting in an
opposite values of the electrostatic potential at some distances 
\cite{tanaka2001chargeinversion}.
The charge compensation parameter of DNA
 phosphate charges, defined as  
\begin{eqnarray}
& \theta(r) = e \int_0^r {\left(-4 \rho_Q(r') +  \rho_+(r') -
  \rho_-(r')   \right) } 
 \nonumber & \\ &
\times 2 \pi r' dr' + N_P  ,&
\end{eqnarray}
accounts for the integrated total charge at distance $r$ away from the DNA axis.
Here $\rho_i(r)$ $(i=Q,+,-)$ are radial ion charge densities per pitch length
and $e>0$ is the elementary charge. 
Data for $\theta$, often called a distance dependent effective DNA
charge, are plotted in Figure \ref{figure_3} for different Spe and salt densities. 

\begin{figure}
   \epsfxsize=8cm
   \epsfysize=8cm
~\hfill\epsfbox{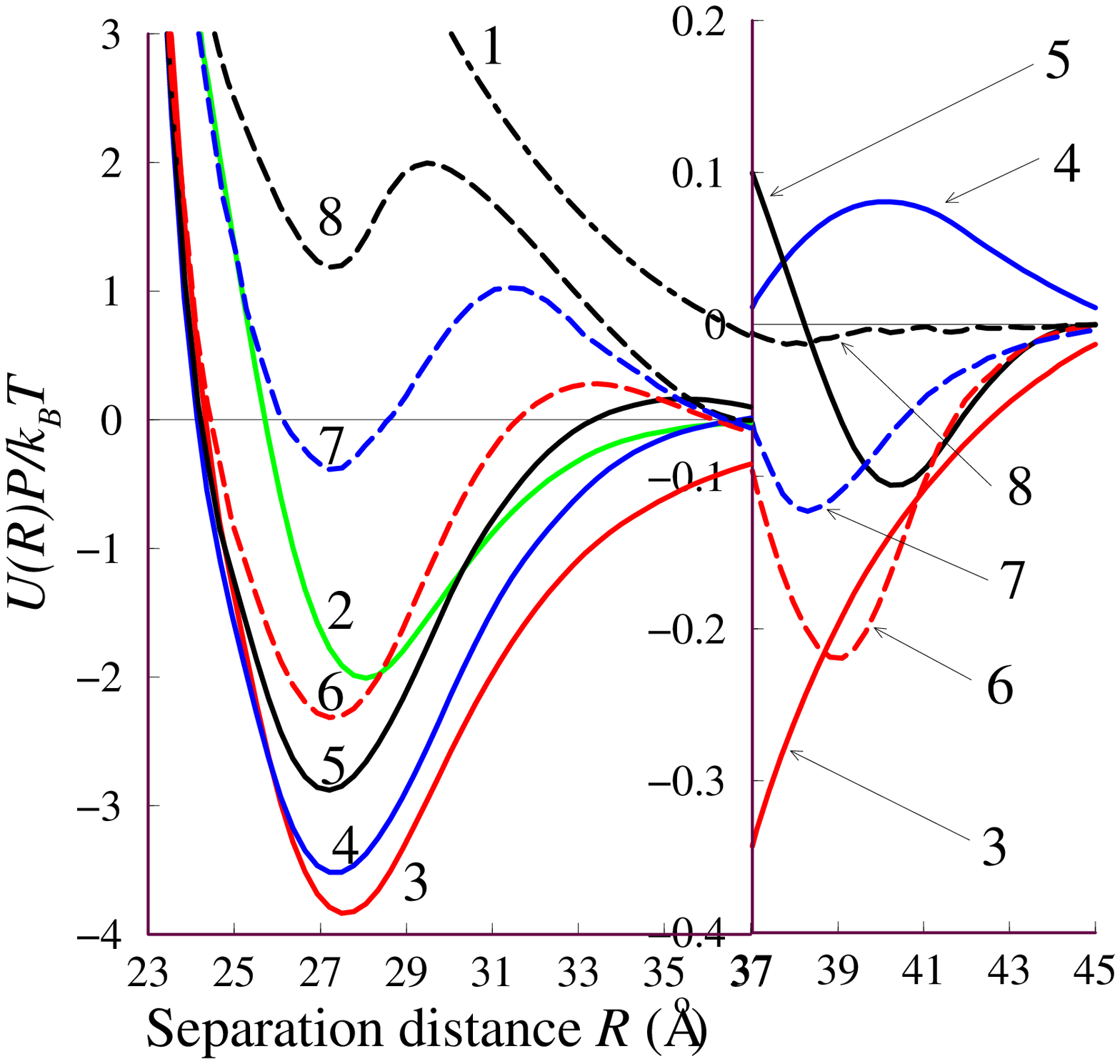}\hfill~
  \caption{(Color online) Effective pair potential  between DNA molecules for $c_s$=25mM
   and $C$=0mM (1), 0.1mM (2), 0.8mM (3), 18mM (4), 65mM (5), 160mM
   (6), 280mM (7), 400mM (8). } 
 \label{figure_5}
\end{figure}

For dense Spe concentrations 
a layered structure appears in $\theta$ due to the bulk charge-density
 oscillations in  strong electrolytes \cite{Attard_review_1996}. 
The simulations reveal qualitatively different types of competition between
the Spe and salt ions on the DNA surface
depending on whether the DNA is overcharged
or not. We find that, at a constant salt density condition and for $C<$1.8mM
(undercharged DNA case), the added Spe ions replace the small
counterions on the DNA surface. 
For Spe concentration $C>$1.8mM (overcharged DNA) 
 i) the influence of salt on the parameter $\theta$
diminishes, in other words, once the DNA is
overcharged, it reluctantly responds to added salt, 
ii) the number of condensed counterions remains constant,
the total adsorbed ionic charge in DNA grooves remains
constant. It is only the
strand  ionic charge which increases
gradually as more Spe is added to solution. Thus, the DNA
overcharging emerges mainly due to the excess Spe charge adsorbed
on the DNA
phosphate strands. 
The rest of this letter is devoted to the question, how the overcharged DNA
and the alternating charge layers affect the 
interaction between {\it two} DNA molecules. To reveal this impact, we
calculate the total effective pair interaction potential $U(R)$ per
unit length for a given bulk salt concentration
 $c_s$ and different Spe concentrations $C$. The quantity $U(R)$
 is gained by integrating the distance-resolved interaction force
 averaged over all microion configurations \cite{allah}. 
Results are shown in Figure~\ref{figure_5}. 
It is seen that even a small trace
of spermine ions - well below the overcharging threshold - induces an attraction between the DNA molecules,
except at very close
distances, see curve (2) for $C$=0.1mM. This attraction has mainly
a pure electrostatic origin and arises due to charge correlations
in the electrolyte. 
For increasing $C$, this (first) minimum is getting deeper and is achieving a maximal
depth at the overcharging concentration $C \approx 1.8$mM. There the minimum
is mainly resulting from entropic forces, i.e.\ from 
layering in the Spe number density. Further increasing $C$, again reduces
the depth of the first minimum. The position of the first minimum, on
the other hand,
hardly depends on $C$. The combined Spe-layering around the pair
of DNA molecules induces a second minimum at larger
separations as revealed in the enlarging inset of Figure 3.
This minimum is of electrostatic origin and occurs for $C\gtrsim 65$mM.
Again the depth of the second minimum increases and decreases with $C$. 
At intermediate Spe concentrations,
we are thus confronted with a double minimum potential which is induced
by layering. 
Bearing in mind that the
 potentials in
Figure~\ref{figure_5} are scaled for one DNA pitch length,
very long DNA molecules may easily collapse into the minimum of curve
(2). This implies  that  DNA aggregation can take
place well {\it below} overcharging  Spe concentrations.
\noindent 

The characteristic double-minimum structure of the interaction potential
$U(R)$ will give rise to unusual phase behavior. We have calculated the
 phase diagram of a columnar DNA assembly on the basis of our
simulated effective pair interactions. We assume
that the DNA molecules are parallel along a certain length $\ell$.
This length is an additional parameter 
which we fix to be $\ell =20 \times P$. We comment on the dependence
of the phase diagram on $\ell$ later. 
The assembly of parallel DNA can be considered as a two-dimensional
many-body system interacting via $U(R) \times \ell$ and being
characterized by a DNA particle number density $\rho$. 
We calculated the free energies of the fluid
and solid phases by using different techniques outlined below and perform the
traditional Maxwell double tangent to identify the coexisting densities.

The free energy of dilute {\it fluid} phase is approximated by the 
two-dimensional perturbation theory according to 
Weeks, Chandler and Anderson \cite{wca1979}. The total
potential is split into a repulsive part $U_r(R)$ and a an attractive
part $U_a(R)$. The former
is identical to $U(R)$ but truncated and shifted towards zero
at the first minimum at $R=R_{min}$. This repulsive potential is then mapped
onto that of effective hard disks of diameter $\sigma_{eff}$ using the  Barker-Henderson 
formula \cite{evans_leshouse} $\sigma_{eff} = \sigma +
\int_{\sigma}^{R_{min}}{(1-\exp{(-\frac{U_r(R)}{k_BT}) }) }dR$.
Here the cross-section diameter for the DNA molecule is $\sigma=20\AA$.  
The total Helmholtz free energy involves that of a hard disk fluid with
effective area fraction $\eta=\frac{\pi \rho \sigma_{eff}^2}{4}$
and a mean-field correction which we simply model as
$\pi \rho^2 \int_{\sigma}^{\infty}{ \frac{U_a(R)}{k_BT} R
dR}$. For the former quantity analytical expressions are available
\cite{harddisks}. 
The free energy of the {\it solid} phase, on the other hand,  is
calculated by a lattice sum 
assuming a two-dimensional triangular lattice as a possible candidate
structure. The lattice constant 
is directly related to the DNA number density $\rho$.
\begin{figure}
   \epsfxsize=8cm 
   \epsfysize=8cm 
~\hfill\epsfbox{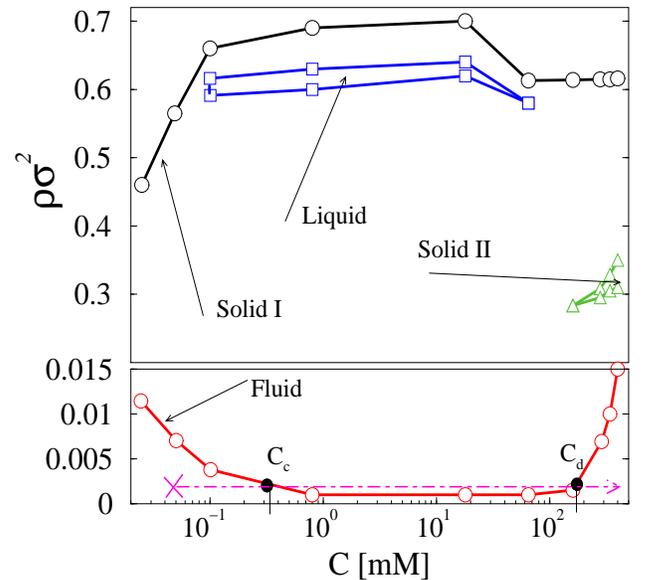}\hfill~
  \caption{ (Color online) Coexisting DNA densities at different Spe
   concentrations $C$ and for $c_s$=25mM. 
The stable phases found are fluid, liquid and two triangular
   crystals with different 
lattices constants (solid I and solid II). All phase transition
   between these phases are of 
   first order. For the sake of better resolution at smaller DNA
   densities, the {\it y}-axis is expanded below $\rho \sigma^2<0.015$.}   
 \label{figure_8}
\end{figure}
The resulting phase diagram with the coexisting DNA densities is shown
in Figure~\ref{figure_8}  for  
the whole range of Spe concentrations $C$. At low $C$ there is  a
strong first-order fluid-crystal 
phase transition. Increasing $C$ widens the coexistence region
considerably due to the 
increasing attractions. Above a threshold concentration of $C\approx
0.1$mM there is enough 
attraction to stabilize a liquid phase of high DNA density. The stability of the liquid
ceases at  
$C\approx 65$mM.
At even higher Spe concentrations a second crystal with a considerable
smaller 
lattice constant than that of the high-density solid
emerges. We call 
this novel phase a mesocrystal since its density is intermediate
between that 
of the fluid and the other almost closed-packed solid.

Another implication of the phase diagram is the condensation and
subsequent redissolution. 
Let us start at small $C$ with a dilute DNA solution (see the cross in
Figure~\ref{figure_8}) and increase $C$ (dot-dashed line in
Figure~\ref{figure_8}). We keep 
$\rho$ fixed to $\rho \sigma^2=0.002$ which
corresponds to a typical DNA concentration of 1mg/ml DNA.  
First the fluid-liquid coexistence
line is hit, which implies that the system will split into a low
density fluid and a high density 
liquid which is the condensation transition. At much higher $C$ the
coexistence line is touched again 
and the system redissolutes back into the dilute fluid phase. The 
corresponding threshold
concentrations of the condensation and redissolution are in the range 
$C_c\approx 0.3$mM and $C_d\approx 165$mM and agree well with the
experimental findings \cite{raspaud1999,pelta1996}. 

Let us finally comment on the dependence of the phase diagram
on the DNA length $\ell$.
Since $\ell$ is a  prefactor of the effective potential
it plays formaly the role of an inverse system temperature.
We have explored the phase behaviour for  smaller DNA segment lengths
of $\ell=5P$ and for larger $\ell=100P$. As a result,  the stability
of liquid pocket depends sensibly on $\ell$: it disappers
completely for small $\ell$ but extends towards larger $C$ for
larger $\ell$. The second feature concerns the fluid coexistence density:
it shifts to considerably higher values for smaller $\ell$.
Hence, condensation and redissolution is prohibited for small
DNA-segment lengths This is in line  with the experiments of Ref.
\cite{tripletDNAgoobes2002} where a threshold value of $\ell\approx 15P$ for the minimal
length $\ell$ required for condensation is reported.

In conclusion, we have calculated the influence of tetravalent counterions 
on the effective interaction and the phase diagram
of columnar DNA assemblies by primitive-model-type computer simulations
and statistical theories. We find that a small concentration of tetravalent
counterions induces DNA condensation. The layering of the strongly coupled
tetravalent counterions to the DNA strands yields an oscillatory
effective interaction potential with a double minimum structure at intermediate
counterion concentrations. This explains the redissolution transition
and triggers a novel stable mesosolid. Our threshold concentrations are 
in good agreement with experimental data.

Future work should address the phase behaviour of DNA solutions which are
 polydisperse in their length
 on the basis of the effective interaction found in this paper.
Furthermore it would be interesting to explore the stability of a
 cholesteric phase 
\cite{saminathan,pelta1996,longtriplex1999saminathan,solis2001reentrance,solis2002}
based on  an effective interaction
which incorporates the relative orientation of the two DNA molecules.

Acknowledgments: We acknowledge discussions with R. Blossey.
Financial support from DFG within LO 418/9
is acknowledged.


\begin{references}
\bibitem{footnote}  Here, ions with the same valency are treated as
indistinguishable. For example, counterions that neutralize the DNA
charge  are indistinguishable from salt counterions.
\end{references}
\end{document}